\documentclass[11pt]{article}
\usepackage{a4}
\usepackage{amsmath}
\usepackage{amsfonts}     
\usepackage{epsfig}
\newcommand{\be}{\begin{equation}}
\newcommand{\ee}{\end{equation}}
\newcommand{\bea}{\begin{eqnarray}}
\newcommand{\eea}{\end{eqnarray}}

\renewcommand{\d}{\mathrm{d}}

\DeclareMathSymbol{\mg}{\mathrel}{symbols}{"1D}

\newcommand{\ga}{\alpha}
\newcommand{\gb}{\beta}
\renewcommand{\gg}{\gamma}
\newcommand{\gd}{\delta}

\newcommand{\gf}{\phi}

\newcommand{\gx}{\xi}
\newcommand{\gm}{\mu}
\newcommand{\gn}{\nu}

\newcommand{\gl}{\lambda}
\newcommand{\gr}{\rho}

\newcommand{\gs}{\sigma}

\newcommand{\go}{\omega}
\newcommand{\gz}{\zeta}
\newcommand{\gp}{\pi}

\newcommand{\get}{\eta}

\newcommand{\gG}{\Gamma}
\newcommand{\gD}{\Delta}
\newcommand{\gF}{\Phi}

\newcommand{\gS}{\Sigma}

\newcommand{\cD}{{\cal D}}

\newcommand{\cL}{{\cal L}}

\newcommand{\cP}{{\cal P}}

\newcommand{\cR}{{\cal R}}

\newcommand{\tg}{{\tilde g}}

\newcommand{\slashed}{\hspace{-1.1ex}/}
\newcommand{\Slashed}{\hspace{-1.3ex}/\hspace{.2ex}}

\newcommand{\ra}{\rightarrow}

\renewcommand{\Im}{\text{Im}\ }

\newcommand{\der}{\partial}

\newcommand{\dsp}{\displaystyle}

\newcommand{\labl}[1]{\label{#1}}
\newcommand{\half}{\frac 12 }
\newcommand{\shalf}{{\scriptstyle \half}}

\newcommand{\beq}{\begin{equation}}
\newcommand{\eeq}{\end{equation}}
\newcommand{\barr}{\begin{array}}
\newcommand{\earr}{\end{array}}
\newcommand{\equ}[1]{\begin{gather} #1 \end{gather}}
\newcommand{\equa}[1]{\begin{align} #1 \end{align}}

\newcommand{\tabu}[2]{\begin{tabular}{#1} #2 \end{tabular}}
\newcommand{\arry}[2]{\begin{array}{#1} #2 \end{array}}

\newcommand{\mtrx}[1]{\begin{matrix} #1 \end{matrix}}
\newcommand{\pmtrx}[1]{\begin{pmatrix} #1 \end{pmatrix}}

\newcommand{\non}{\nonumber}
\newcounter{oldcounter}

\newcommand{\bgl}{{\bar\lambda}}

\newcommand{\bgz}{{\bar\zeta}}

\newcommand{\Natr}{\mathbb{N}}
\newcommand{\Intr}{\mathbb{Z}}

\if@twoside\oddsidemargin25mm
\else
\begin{document}

\begin{flushright} 
BONN-TH-34-01 \\
hep-th/0108184
\end{flushright} 
\vskip 2 cm
\begin{center}
{\Large {\bf Gauge corrections and FI-term in 5D KK theories
}}
\\[0pt]
\bigskip
\bigskip {\large
{\bf D.M. Ghilencea, \footnote{
{{ {\ {\ {\ E-mail: dumitru@th.physik.uni-bonn.de}}}}}} }
{\bf S. Groot Nibbelink, \footnote{
{{ {\ {\ {\ E-mail: nibblink@th.physik.uni-bonn.de}}}}}} } 
{\bf H.P. Nilles \footnote{
{{ {\ {\ {\ E-mail: nilles@th.physik.uni-bonn.de}}}}}} }
\bigskip }\\[0pt]
\vspace{0.23cm}
{\it Physikalisches Institut der Universitat Bonn,} \\
{\it Nussallee 12, 53115 Bonn, Germany.}\\
\bigskip
\vspace{3.4cm} Abstract
\end{center}
In the context of a five dimensional N=1 Kaluza Klein model
compactified on $S_1/\Intr_2\times \Intr_2'$ we compute the
one-loop gauge corrections to the self energy of the
(zero-mode) scalar field. 
The  result is quadratically divergent due
to the appearance of a Fayet-Iliopoulos term.

\newpage

\section{Introduction}
There has recently been growing interest in the phenomenological
aspects of large additional (compact) space dimensions in the 
context of Kaluza-Klein models. 
Various models in this direction have been built 
and their  starting point is in general the assumption 
of the existence of a 5 dimensional N=1 supersymmetric model  
compactified on $M^4 \times S^1/\Intr_2$  or on
$M^4 \times S^1/\Intr_2\times \Intr_2'$ 
\cite{Antoniadis,Delgado,Hall,Barbieri,Quiros}.

While the details of these models are rather involved and model
dependent, an interesting generic feature emerges, that
the potential and the Higgs mass in {\it one-loop} order
are ultraviolet insensitive as far as Yukawa contribution is 
concerned. It is not clear to what extent
this property may survive to higher orders, since
the (one loop) couplings of the 4 dimensional theory have in general a 
dependence on the high scale worse than that
(of logarithmic type) of the minimal supersymmetric standard model 
(MSSM). This dependence
may in turn be re-introduced in  the expressions of the
potential and the  Higgs mass beyond one-loop order. 
Therefore high scale sensitivity may be restored
via the couplings of the theory due to the extra dimension
affecting their ``running''. This sensitivity
may be further affected by the 
non-renormalizable character of these models.

Of this class of Kaluza Klein models, our attention was drawn 
to that of reference \cite{Barbieri} which is a very interesting 
construction 
providing an extension of the standard model to a 5 dimensional N=1
supersymmetric theory compactified on an orbifold 
$M^4 \times S^1/\Intr_2\times \Intr_2'$. All Standard Model (SM) 
states have associated Kaluza Klein states with respect to
the extra dimension. Even though the 5 dimensional model
is supersymmetric (before compactification), it has the  distinctive feature 
that its low energy particle spectrum  
contains only one light Higgs doublet, that corresponding to the SM.
In this respect the model is more similar to the standard model than 
to its minimal supersymmetric extension (MSSM). 

It is well known that in the SM model the Higgs sector suffers from an  
extreme fine-tuning due to the quadratic divergence of the mass 
parameter in the Higgs potential. In the MSSM model all quadratic 
divergences are however absent. The reason for this is two-fold: 
the model is  supersymmetric and  the only  quadratic divergent term is the 
Fayet-Iliopoulos (FI)-term\footnote{Here we disregard the renormalization of 
the vacuum energy or cosmological constant term, as it is not clear 
to what extent it fits in a theory that does not take quantum gravity 
into account.}.  However, this divergence is proportional 
to the sum of the hyper-charges of all massless complex scalars. 
In the MSSM this sum is zero, because it is equal to the sum of 
hyper charges of the chiral fermions by supersymmetry. This vanishes 
so that the mixed gravitational-gauge anomaly does not arise. Hence, 
the MSSM does not contain quadratic divergences. 

From this perspective, 
one would like to address the situation of  the model \cite{Barbieri}, 
since it apparently has supersymmetric 
features as it is obtained by compactifying a $N =1$ supersymmetric 
theory in 5 dimensions on the orbifold $S^1/\Intr_2\times \Intr_2'$. 
As a result the Kaluza-Klein (KK) towers fall into multiplets of this 
supersymmetry.  However,  the low energy spectrum 
(of massless states) which is precisely that of the standard model 
has only one Higgs state.  
It is thus not clear to what
extent the  properties of $N =1$ in 5 dimensions 
manifest themselves after the compactification, to protect the 
model from divergences.

Not all properties of the initial $N=1$ theory on 
5 dimensional Minkowski space remain after compactification on an 
orbifold. For example, this  compactification makes it possible 
that a chiral spectrum exists in the effective 4 dimensional theory, 
while the uncompactified version is necessarily non-chiral. In this way
not only can a chiral spectrum arise, but also the possibility of 
anomalies is opened up. In an accompanying paper \cite{SHP} anomalies 
in such theories will be investigated. Not surprisingly, it is found that 
only the exactly massless chiral fermions in the effective 4 dimensional 
theory can contribute to it. This is of course in perfect agreement 
with the index theorems that are behind the anomalies 
\cite{Alvarez,Coleman,Alvarez2}. 

The renormalization of the FI-term, the 
tadpole of the auxiliary field in the vector multiplet, only depends 
on massless states at one loop \cite{Fischler} and is therefore very similar. 
However, for the calculation of the FI contribution we have to rely heavily on 
a regularization prescription. Each scalar mode gives a quadratically 
divergent contribution. 

In the present paper we investigate the possible role of a FI-term
in the model of ref.\ \cite{Barbieri}. Therefore we discuss $N =1$ 
supersymmetric gauge theory coupled to a hypermultiplet in
5 dimensions and its compactification on the orbifold
$S^1/\Intr_2 \times \Intr_2'$. We observe that a FI-term at
the boundary is allowed by all the symmetries of the theory.
We then compute the one loop contribution to the FI-terms using the
method set up in ref.\ \cite{sgn}. We find a quadratic divergence
of the FI-term, which in turn induces a quadratic divergence for the
Higgs mass term.

\section{Five-dimensional vector and hyper multiplet}
\label{EffectiveAction}

In this section we give the 5 dimensional $N =1$ action that 
will be the basis of our later discussion and describe its 
compactification on $M^4 \times S^1/\Intr_2\times \Intr_2'$. 
Before that we briefly motivate our gauge choice 
for the Abelian gauge field. 

The Lagrangian for a 5 dimensional Abelian gauge field $A_M$ 
with field strength $F_{MN}$ reads 
\equ{
\cL_{G} = - \frac 14 F_{MN} F^{MN}.
\labl{Glagr}
}
Since in the 5 dimensional theory the 5th dimension is compact 
while the others are not, the theory has only 4 dimensional Lorentz 
invariance. This implies that also for the gauge fixing term 5 
dimensional Lorentz invariance may be broken. We take 
\equ{
\cL_{G.F.} = - \half 
\left( \ga_4 \, \der_\gm A^\gm + \ga_5 \, \der_5 A^5 \right)^2,
\labl{GFlagr}
}
with $\ga_4$ and $\ga_5$ real gauge fixing parameters. (If we 
take $\ga_4 = \ga_5$ we recover the original 5 dimensional 
gauge fixing condition.) These gauge 
fixing terms can be obtained by integrating the gauge fixing 
$ \ga_4 \, \der_\gm A^\gm + \ga_5 \, \der_5 A^5 = \go$ 
in the path integral over a Gaussian distribution 
$\int \cD \go \exp{i \int \d^5 x \shalf \go^2}$. As usual,  
for an Abelian theory with such a gauge fixing term, the ghost 
sector decouples from the physical part of the theory. 
In momentum space this leads to the form
\equ{
\cL_{G\, tot} = - \half \pmtrx{ A_\gm^* & A_5^* }
\pmtrx{
p^2 \get^{\gm \gn} - ( 1 - \ga_4^2) p^\gm p^\gn & 
- (1 - \ga_4 \ga_5) p^\gm p^5 
\\[1ex]
- (1 - \ga_4 \ga_5) p^5 p^\gm & 
p^2 - ( 1 - \ga_5^2) p_5^2
}
\pmtrx{A_\gn \\ A_5},
}
with $p^2 = p_4^2 + p_5^2$ is the sum of the 4 and 5 dimensional 
momentum squared. Notice that if we take $\ga_5 \ga_4 = 1$, we 
find that the gauge field $A_\gm$ decouples from the scalar $A_5$. 
In addition we set $\ga_5 = 1$, thus the propagator for $A_5$ is
simply proportional to $1/p^2$. In this case $\ga_4 = \ga_5 = 1$, we 
thus obtained the (generalized) Feynman gauge. 
For the one-loop calculation that is performed later, this gauge will be 
used for simplicity. 

The $N = 1$ supersymmetric 5 dimensional Lagrangian 
$\cL = \cL_V + \cL_H$ 
describes a vector and a hyper multiplet. 
The on-shell form of this Lagrangian can be 
found in \cite{Gunaydin,Pomarol,Delgado}; 
the off-shell formulation is 
given in \cite{Mirabelli} (for the vector multiplet) and 
\cite{deWit,Zucker}. 
The components of the vector multiplet 
$V = (A_M, \gl_i, \gF, D^a)$ are, apart from the vector field $A_M$, 
a symplectic 
Majorana gaugino $\gl$, a real scalar field $\gF$ and an $SU_R(2)$ 
iso-triplet of auxiliary scalars $D^a$. Also the gaugino transforms 
as a doublet under $SU_R(2)$. Including the gauge-fixing term in 
the Feynman gauge, their action is given by 
\equ{
\cL_V = 
- \frac 14 F_{MN} ^2 - \half (\der_M A^M)^2 
- \bgl \der\slashed \gl - \half (\der_M \gF)^2 + \half (D^a)^2.
\labl{lagrV}
}
The hyper multiplet $H = (h^i_\ga, \gz_\ga, F^i_\ga)$ consists of 
a bi-doublet of $SU(2)\times SU_R(2)$ 
scalars $h^i_\ga$, a hyperino $\gz_\ga$ and a bi-doublet 
auxiliary scalars $F^i_a$. 
This hyper multiplet couples to $A_M$ as it caries hyper charge; 
the derivatives have become covariant derivatives with gauge 
coupling constant $\tg$ in 5 dimensions. This leads to the Lagrangian 
\equ{
\cL_{H} = - | D_M h^i_\ga |^2 - \bgz^\ga (D\Slashed + \tg \gF) \gz_\ga 
+ |F^i_\ga|^2 - ( i \sqrt 2\tg\, h^{\dag \ga}_i \bgl^i \gz_\ga + \text{h.c.}) 
\non \\ 
- \tg^2\,h^{\dag \ga}_i \gF^2 h^i_\ga 
- \tg D^a h^{\dag \ga}_i \gs_a h^i_\ga. 
\labl{lagrH} 
}

Let us consider now  the situation that arises when these fields are
placed on the  
orbifold $S^1/\Intr_2 \times \Intr_2'$. This orbifold is defined by the 
periodicity $x_5 \ra x_5 + 2 \gp R$ and the two parities 
$x_5 \ra - x_5$ and $x_5 \ra \gp R - x_5$. 
By performing similar analyses 
as discussed in \cite{Bergshoeff,SHP}, one finds that both the gaugino 
and the hyperino have to transform under the parities in a non-trivial 
way. An allowed (though not unique) choice is given by 
\equ{
\mtrx{
\gl(-x_5) = a_R\gg^5\, \gl(x_5), & 
\gl(\gp R - x_5) = - a_R\gg^5\, \gl(x_5), 
\\[1ex]
\gz(-x_5) = a\gg^5\, \gz(x_5), & 
\gz(\gp R - x_5) = - a\gg^5\, \gz(x_5),
}
}
where $a$ and $a_R$ are elements of the algebras of $SU(2)$ 
and $SU_R(2)$, respectively.
This means, for example, that $a_R = \vec a_R \vec \gs$, with 
$\vec \gs$ the vector of Pauli matrices.  
In ref.\ \cite{Barbieri} this is done with $a = a_R = \gs_3$.  
One then finds the parity assignments for the Kaluza-Klein modes
(see also Appendix \ref{modepoleS22})
\begin{center}
\tabu{l | l | l | l | l | l | l | l | l | l }{
fields & 
$h^\ga_{n\,\ga}$ & $h^\ga_{n-\ga}$ & $\gz^\ga_n$ & $A_\gm^n$ & 
$A_5^n$ & $\gl_i^n$ & $\gF_n$ & $D^\parallel_n$ & 
$\vec{D}^\perp_n$ 
\\ \hline 
parities & $--$ & $++$ & $\ga-\!\ga$ & $++$ & $--$ & $i-\! i$ & 
$--$ & $++$ & $--$  
\\
modes $n$ & $ \geq 1$ & $\geq 0$ & $\geq 0$ & $\geq 0$ & $\geq 1$ & 
$\geq 0$ & $\geq 1$ & $\geq 0$ & $\geq 1$
}
\end{center}
with $\ga, i = \pm$. 
Of the three auxiliary scalars, that form a triplet $\vec D$ 
under $SU_R(2)$ of the $N = 1$ vector multiplet, 
two are odd under both parities 
$\vec D^\perp = (1 - \vec a_R \vec a_R^T) \vec D$, while the 
other one $D^\parallel = \vec a_R^T \vec D $ is even. 
The resulting Feynman rules for these KK fields can be 
derived from the Lagrangians  by using products rules for 
the mode functions and their orthonormality properties 
\eqref{orthS22}. 

\section{The Fayet-Iliopoulos term}

It is well-known that in a (unbroken) supersymmetric field theory in 
4 dimensions the FI-term is either quadraticly divergent or vanishes 
at one loop.  The diagram of the FI-contribution to the selfenergy of a 
scalar\footnote{This can be the zero mode or any of the KK excitations.} 
is given by,
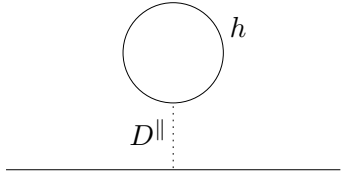
\begin{center}
\begin{picture}(0,0)%
\includegraphics{FI.pstex}%
\end{picture}%
\setlength{\unitlength}{2763sp}%
\begingroup\makeatletter\ifx\SetFigFont\undefined%
\gdef\SetFigFont#1#2#3#4#5{%
  \reset@font\fontsize{#1}{#2pt}%
  \fontfamily{#3}\fontseries{#4}\fontshape{#5}%
  \selectfont}%
\fi\endgroup%
\begin{picture}(3024,1519)(2089,-5773)
\put(4126,-4561){\makebox(0,0)[lb]{\smash{\SetFigFont{11}{13.2}{\familydefault}{\mddefault}{\updefault}
$h$
}}}
\put(3226,-5536){\makebox(0,0)[lb]{\smash{\SetFigFont{11}{13.2}{\familydefault}{\mddefault}{\updefault}
$D^\parallel$
}}}
\end{picture}

\end{center}
where the dotted line corresponds to the auxiliary field $D^\parallel$ 
of the Abelian gauge multiplet in 4 dimensions. 
We investigate what happens to the FI-term in the effective field theory 
coming from 5 dimensions with a mass spectrum of the complex scalars 
of the hyper multiplet on $S^1/\Intr_2 \times \Intr_2'$,  as discussed in 
appendix \ref{modepoleS22}. We take the charges of these scalars 
such that $q^{++}_n = - q^{--}_n = 1$ because they belong to complex
conjugate representations. Formally, the expression for 
the one loop contribution to the FI term reads 
\equ{
\gx^{} = \sum_{n, \ga} \,g q^{\ga\ga}_n \int \frac{\d^4 p_4}{(2\gp)^4} 
\frac {1}{p_4^2 + ({m_n^{\ga\ga}})^2 + m^2},
\labl{FIoneloopKK}
} 
where $m_n^{\ga\ga} = 2n/R$ and the sum for $\ga = +$ is over 
$n \geq 0$, while for $\ga = -$ over $n > 0$. 
Here the 4 dimensional gauge coupling $g$ is related to the 5 dimensional 
gauge coupling $\tg$ by $g^2 = \frac 4{\gp R} \tg^2$. 
We have introduced 
an IR regulator mass $m$, which is needed to turn the sum into 
a contour integral by complex function analysis \cite{sgn}. 
To obtain this expression we made the following 
observations. Because the components of the vector multiplet are all 
5 dimensional fields, for any $n$-point function we have 5 dimensional 
momentum and parity conservation. This means in particular that there 
is no momentum flow into the tadpole and that the parity of the 
tadpole is $++$. Therefore only the zero mode of $D^\parallel$ is 
relevant here. Since only scalars $h$ run around in the 
loop, their parity is either $++$ or $--$. The vertices are then 
$\gd_{nn'} + \gd_{n0}\gd_{n'0}$ and 
$-\gd_{nn'} $ for $++$ and $--$ parity scalars, respectively,  
because of the orthonormality relation \eqref{orthS22} and their 
charges. Similarly, for the propagators we have 
\equ{
\gD^{++}_{n n'} = \frac 1{p_4^2 +(2n/R)^2} 
(\gd_{nn'} - \half \gd_{n0}\gd_{n'0}), 
\qquad
\gD^{--}_{n n'} = \frac 1{p_4^2 +(2n/R)^2} 
\gd_{nn'}. 
} 
Notice that the different normalization of the propagator 
the massless mode is compensated by the normalization of the 
vertex, so that we can simply sum all contributions with the 
same weight. 

As the expression \eqref{FIoneloopKK} contains a potentially divergent 
sum and integral, it can only be defined unambiguously using some 
regularization. We use dimensional regularization for both the integral and 
the sum, as was discussed in ref.\ \cite{sgn}. The relevant pole functions 
$\cP^{++}$ and $\cP^{--}$ for the momentum spectra of $\gf_n^{++}$ 
and $\gf_n^{--}$ are given in eq.\ \eqref{polefunS22} of appendix 
\ref{poleS22}. 
According to the dimensional regularization procedure,  we obtain
\equ{
\gx^{} = g\, \frac {1}{2\gp i}\,  \int \frac{\d^{D_4} p_4}{(2\gp)^{D_4}} 
\int_\ominus \d^{D_5} p_5
\left\{
\frac{\cP^{++}(p_5)}{p_4^2 + p_5^2 + m^2} - 
\frac{\cP^{--}(p_5)}{p_4^2 + p_5^2 + m^2}
\right\}.
}
Since usual the complex dimensions $D_4$ and $D_5$ are chosen such that 
the whole expression is convergent and then analytically continued. 
This procedure is summarized in appendix \ref{poleS22}. 
Substituting the expressions of the pole functions 
\eqref{polefunS22}, gives exactly the same result 
as the regulated FI term for one massless complex scalar:
\equ{
\gx^{} = g\, \frac {1}{2\gp i}\,  \int \frac{\d^{D_4} p_4}{(2\gp)^{D_4}} 
\int_\ominus \d^{D_5} p_5\, 
\frac 1{p_5} 
\frac {1}{p_4^2 + p_5^2 + m^2} 
= g\, 
\int \frac{\d^{D_4} p_4}{(2\gp)^{D_4}} \frac {1}{p_4^2 + m^2}.  
\labl{FItermzero}
}
Since the result (\ref{FItermzero})  behaves like
a zero-mode particle contribution we can safely take 
$D_5 =1$ and remove the contour integration, 
to obtain exactly the 4 dimensional expression.\footnote{If the 
hyper multiplet is a doublet under $SU_L(2)$ this result is multiplied 
by 2.} This shows that the FI term at one loop is simply given by the 
contribution of the  massless complex scalar, putting $m \ra 0$. 
This result holds for any finite $R$, since it is independent 
of the radius $R$ of the compact dimension. Therefore, we conclude 
that it is also true in the limit $R \ra \infty$. This signals
that the boundary condition of  the 
orbifolding is not removed in this decompactification limit. 

We have performed a similar calculation for the case in which we have 
complex scalar states that have odd wave functions ($\gf^{+-}_n$ and 
$\gf^{-+}_n$). However, since in that case all fermionic states 
have nonvanishing masses, their contribution to a FI term is zero. 

For this Fayet-Iliopoulos contribution, an auxiliary field tadpole counter 
term has to be introduced. 
Such a counter term of course has to be consistent with the symmetries 
of the theory. On both branes we have at most $N = 1$ supersymmetry 
because the other supersymmetry has the wrong parity to exist on that 
brane. Therefore, on the branes $D$-terms can be added for the 
auxiliary fields that do not vanish. As can be seen from the table 
in section \ref{EffectiveAction} the only auxiliary field that exists 
on either brane is $D^\parallel$.

\section{Other gauge  corrections to the self energy}

In the previous section we have only focused on the correction to 
the tadpole of the component of the auxiliary field $D^\parallel$ of 
the vector multiplet. This result gives in turn a contribution 
to the renormalization of the scalar masses. In this section we 
calculate other gauge coupling corrections to the self energy of 
scalar fields. Here we restrict ourselves to the corrections to
the self energy of the zero mode scalars only. 
The relevant diagrams are given below
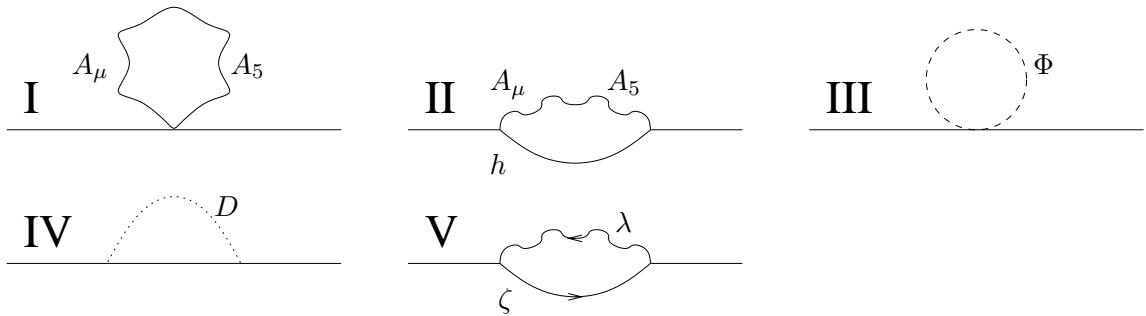
\begin{center}
\begin{picture}(0,0)%
\includegraphics{GaugeCon.pstex}%
\end{picture}%
\setlength{\unitlength}{2763sp}%
\begingroup\makeatletter\ifx\SetFigFont\undefined%
\gdef\SetFigFont#1#2#3#4#5{%
  \reset@font\fontsize{#1}{#2pt}%
  \fontfamily{#3}\fontseries{#4}\fontshape{#5}%
  \selectfont}%
\fi\endgroup%
\begin{picture}(10224,2754)(2089,-7403)
\put(7501,-5386){\makebox(0,0)[lb]{\smash{\SetFigFont{11}{13.2}{\familydefault}{\mddefault}{\updefault}
$A_5$
}}}
\put(7576,-6661){\makebox(0,0)[lb]{\smash{\SetFigFont{11}{13.2}{\familydefault}{\mddefault}{\updefault}
$\gl$
}}}
\put(3976,-6511){\makebox(0,0)[lb]{\smash{\SetFigFont{11}{13.2}{\familydefault}{\mddefault}{\updefault}
$D$
}}}
\put(4126,-5236){\makebox(0,0)[lb]{\smash{\SetFigFont{11}{13.2}{\familydefault}{\mddefault}{\updefault}
$A_5$
}}}
\put(2701,-5236){\makebox(0,0)[lb]{\smash{\SetFigFont{11}{13.2}{\familydefault}{\mddefault}{\updefault}
$A_\gm$
}}}
\put(6526,-7336){\makebox(0,0)[lb]{\smash{\SetFigFont{11}{13.2}{\familydefault}{\mddefault}{\updefault}
$\gz$
}}}
\put(6451,-6136){\makebox(0,0)[lb]{\smash{\SetFigFont{11}{13.2}{\familydefault}{\mddefault}{\updefault}
$h$
}}}
\put(11326,-5236){\makebox(0,0)[lb]{\smash{\SetFigFont{11}{13.2}{\familydefault}{\mddefault}{\updefault}
$\gF$
}}}
\put(6451,-5386){\makebox(0,0)[lb]{\smash{\SetFigFont{11}{13.2}{\familydefault}{\mddefault}{\updefault}
$A_\gm$
}}}
\end{picture}

\end{center}
Here a wavy line stands for a gauge field ($A_\gm, A_5$), 
a wavy line with an arrow  a gaugino $\gl$, 
a line with an arrow a hyperino $\gz$, 
a dashed line a real scalar $\gF$ and a dotted line is an 
auxiliary field $D^a$. On the orbifold $S^1/\Intr_2 \times \Intr_2'$ 
these fields are classified as even or odd under both parities, 
see table in section \ref{EffectiveAction}. 

We use dimensional reduction \cite{Capper} to treat the fermions, 
i.e.\ the fermionic traces are computed in 5 dimensions, 
the resulting sum and integrals are dimensionally regulated \cite{sgn}. 
We denote the gauge correction (except the FI tadpole) 
to the self energy by $\gS_G$ and write  the integrant as the sum
of five terms, corresponding each to the diagrams given above
\equ{
-i \gS_G =  
\frac {-1}{2\gp i}\,  \int \frac{\d^{D_4} p_4}{(2\gp)^{D_4}}  
\int_\ominus \d^{D_5} p_5
\left( 
\text{I} +  \text{I} + \text{II} + \text{III} + \text{IV} + \text{V}
\right). 
}
Using the Feynman rules resulting from the 
Lagrangians \eqref{lagrV}, \eqref{lagrH} 
and the table given in section \ref{EffectiveAction} we find
\equ{
\arry{l}{
\arry{ll}{
\text{I~~} = - 4 g^2 {\dsp \frac 1{p^2}} \cP^{++} 
- g^2{\dsp  \frac 1{p^2}} \cP^{--},
&
\text{II~} = +g^2 {\dsp \frac {p_4^2}{(p^2)^2}} \cP^{++} 
+ g^2 {\dsp \frac {p_5^2}{(p^2)^2}} \cP^{--}, 
\\
\text{III} = - g^2 {\dsp \frac 1{p^2}} \cP^{--},  
&
\text{IV} = - g^2 {\dsp \frac 1{p^2}} \cP^{++} 
- 2  g^2 {\dsp \frac 1{p^2}} \cP^{--},  
} \\
\text{\, V~} = 4g^2 \frac 1{p^2} (\cP^{+-} + \cP^{-+} ).
}
}
Here the factor $\half$ is a sign that the fermion states have become 
chiral. The pole-functions \eqref{polefunS22} are used to rewrite 
the sums over the KK momentum as contour integrals.

To identify the 5 dimensional, 4 dimensional and finite contributions
of these diagrams, we can use the expressions for the pole 
functions given in \eqref{polefunS22>} and \eqref{polefunS22<} 
for $\Im p_5 > 0$ and $\Im p_5 < 0$, respectively. The purely 
5 dimensional divergences for $\Im p_5> 0$ (and for $\Im p_5< 0$ 
for the same reason) vanishes:
\equ{
(\text{I} + \ldots + \text{V})_{5D} = 
- \frac{g^2}{2}\frac 1{p^2}   \Bigl( - \frac i2 \gp R \Bigr)
\left[
4 + 1 - \frac {p_4^2}{p^2} - \frac{p_5^2}{p^2} + 1 + 3 -8 
\right] = 0.
}
This cancellation is reminiscent of the full $N = 1$ supersymmetry 
in 5 dimensions in the uncompactified theory. (If one takes the 
more general gauge choice $\ga_4 = \ga_5 \neq 1$, this contribution 
does not vanish anymore.) 

The purely 4 dimensional divergences of these diagrams vanish also. 
For the integrant we obtain 
\equa{
(\text{I} + \ldots + \text{V})_{4D} = &\
- \frac{g^2}2 \frac 1{p^2} 
\Bigl(
4 - 1 - \frac {p_4^2}{p^2} + \frac{p_5^2}{p^2} - 1 + 1 - 2
\Bigr)\frac 1{p_5} 
= - \frac {g^2}2 \frac 1{p^2} \, 
2 \frac{p_5}{p^2}.
}
In the corresponding integral we set $D_5 = 1$, because 
this contribution does not contain an infinite sum. 
Furthermore, it can be written as a contour around the real axis, 
using the reverse steps as in \eqref{contourints}. 
However, since there is no 
pole of $p_5$ on the real axis, this contribution vanishes 
identically.  (The same result is obtained for the more general 
gauge fixing $\ga_4 = \ga_5 \neq 1$.) 

Since we do not find any quadratic divergences here, it is clear 
that it is not possible to cancel the quadratic contribution of 
the FI-term \eqref{FItermzero} to the scalar mass. This 
situation is similar to the case of 
$N=1$ supersymmetric models in 4 dimensions.

Finally, we can identify the finite contributions. For $\Im p_5 > 0$,  
their integrants read 
\equ{
(\text{I} + \ldots + \text{V})_{finite \, >} = 
- \frac {g^2}2 \frac 1{p^2} 
\left[ \Bigl( 4 + 1 - \frac {p_4^2}{p^2} - \frac{p_5^2}{p^2} + 1 +3
\Bigr) (-\gr_-(p_5)) - 8 \gr_+(p_5)
\right]
\non \\
 = 
+ 4 g^2 \frac 1{p^2} \left[ \gr_-(p_5) + \gr_+(p_5) \right].
}
For $\Im p_5 < 0$, we find
\equ{
(\text{I} + \ldots + \text{V})_{finite \,<} = 
- 4 g^2 \frac 1{p^2} \left[ \gr_-(-p_5) + \gr_+(-p_5) \right].
}
Since this is a finite contribution, we remove the regulators: 
$D_5 = 1$ and $D_4 = 4$ and we obtain (using   \eqref{finitepolefunS22})
\equ{
- i \gS_G = i \frac {7 g^2}{16 \gp^4} \Bigl( \frac 2{R} \Bigr)^2  
\gz(3).}

\section{Conclusion}

We have discussed gauge corrections to the mass parameter in the 
scalar potential in the effective 4 dimensional field theory, obtained 
from a $N =1$ supersymmetric field theory in 5 dimensions. 
In particular, we have seen that the tadpole contribution to a 
component of the auxiliary field is quadratically divergent and 
proportional to the sum of massless scalar fields. This result was 
obtained by using dimensional regularization on both Kaluza-Klein 
sum and 4 dimensional momentum integral separately. 
Using the properties of the pole functions 
associated with the orbifold, it is not difficult to identify the 5 and 
4 dimensional divergent contributions and the additional, finite
parts. 
For the tadpole of the auxiliary field it turned out that only a purely 
4 dimensional divergence is left. 

This situation is very similar to that of 4 dimensional supersymmetric 
field theories, where the FI-term gives an identical result. 
This contribution cannot be removed: 
even if we take $R\ra \infty$ we find the same expression due to the 
massless modes. In this sense, the behavior is very similar to 
anomalies, where only the zero-mode fermions contribute
\cite{SHP}.  

Apart from this Fayet-Iliopoulos contribution, we have calculated 
other gauge corrections to the mass parameter in the effective 
potential. We have shown explicitly that both the possible 
5 and 4 dimensional divergences cancel, leaving a finite contribution. 
Therefore, the quadratically divergent FI contribution cannot 
be compensated. Thus, only in models where the sum of the 
hyper charges of the massless complex scalars is zero can  the 
quadratic divergence to the FI-term be absent. 

In the ${S^1/\Intr_2\times \Intr_2'}$ model 
under consideration we have therefore
shown unambiguously that FI-term and Higgs mass are quadratically
divergent at one loop. 
For a somewhat complementary discussion of this issue see 
ref.\ \cite{GNS}.

\section*{Acknowledgments} 

The authors thank A. Dedes for  helpful discussions.
This work is supported by priority grant 1096 of the Deutsche 
Forschungsgemeinschaft and European Commission RTN 
programmes HPRN-CT-2000-00131 and 
HPRN-CT-2000-00148.

\section*{Appendix}
\appendix
\section{Mode functions 
$\boldsymbol {S^1/\Intr_2\times \Intr_2'}$}
\labl{modepoleS22}

The mode functions 
$\gf^{\ga \gb}_n(x_5)$ can now be even or odd under either of 
these $\Intr_2 \times \Intr_2'$ symmetries:
\equ{
\Intr_2: ~
\gf^{\ga\gb}_n(-x_5) = \ga \gf^{\ga\gb}_n(x_5),
\qquad
\Intr_2': ~
\gf^{\ga\gb}_n(\gp R -x_5) = \gb \gf^{\ga\gb}_n(x_5),
\labl{par22}
}
with $\ga,\gb = \pm$ as eigenvalues. 
Their real representations are 
\equ{
\begin{array}{l l l}
\gf_n^{++}(x_5) = \cos \frac{2nx_5}R, &  n \geq 0,
\qquad 
\gf_n^{--}(x_5) = \sin \frac{2nx_5}R, & n > 0,
\\[2ex]
\gf_n^{+-}(x_5) = \cos \frac{(2n+1)x_5}R, &  n \geq 0,
\qquad
\gf_n^{-+}(x_5) = \sin \frac{(2n+1)x_5}R, &  n \geq 0.
\end{array}
\labl{modeS22}
}
From this it is easy to see that under differentiation 
\equ{
\begin{array}{l  l}
\der_5 \gf^{\ga\ga}_n(x_5) ~~= - \ga \frac {2n}R\, \gf^{-\ga-\ga}_n(x_5), 
&  
\der_5^2 \gf^{\ga\ga}_n(x_5) ~~= - \frac{4n^2}{R^2}\,\gf^{\ga\ga}_n(x_5),
\\[2ex]
\der_5 \gf^{\ga-\ga}_n(x_5) = - \ga \frac {2n+1}R\, \gf^{-\ga\ga}_n(x_5), 
& 
\der_5^2 \gf^{\ga-\ga}_n(x_5) = - \frac{(2n+1)^2}{R^2}\,\gf^{\ga-\ga}_n(x_5).
\end{array}
\labl{derS22} 
}
The Laplacian $\der_5^2$ gives rise to the KK masses in the effective 
field theory.   
The orthonormality of the mode functions $\gf^{\ga\gb}_n$ takes the form 
\equ{
\frac 4{\gp R}  \int_0^{\shalf \gp R} \d x_5\,  
\gf^{\ga\gb}_n(x_5) \gf^{\ga'\gb'}_{n'}(x_5) 
= \gd^{\ga\ga'}\gd^{\gb \gb'} 
( \gd_{n n'} + \gd^{\ga+}\gd^{\gb+} \gd_{n0} \gd_{n'0} ).
\labl{orthS22}
}

\section{Dimensional regularization of the orbifold}
\labl{poleS22}

The pole functions, that can be obtained by similar arguments as 
presented in \cite{sgn}, 
\equ{
\cP^{\ga\ga} = \half 
\left(
\frac \ga{p_5} + \frac {\half \gp R}{\tan \half \gp R p_5} 
\right), 
\qquad
\cP^{\ga-\ga} = - \half 
\frac {\half \gp R}{\cot \half \gp R p_5}, 
\labl{polefunS22}
}
can be used to turn sums into contour integrals. For example,
for a convergent sum we have  
\equ{
\sum_{n \geq 0} f(\frac {2n}R) = 
 \frac {-1}{2\gp i} \int_{\rightleftharpoons} \d p_5\, 
\cP^{++}(p_5) f(p_5) = 
 \frac {1}{2\gp i} \int_{\ominus} \d p_5\, 
\cP^{++}(p_5) f(p_5).
\labl{contourints}
}
The contour $\rightleftharpoons$ consists of two lines 
along the real axis (that are infinitesimally near to it) and 
closed at $\pm$ infinity. 
The contour $\ominus$ contains the full complex plane, 
except the real axis, and is  anti-clockwise oriented. 

Combinations of sum and integrals of a function $f(p_4, p_5)$ 
can be regulated by  
\equa{
\int \d^4 p_4\, 
\sum_{n\in \Natr} \, f(p_4, m_n)  \ra & \  
\frac {1}{2\gp i}\, 
\int_\ominus \d^{D_5} p_5\,  \int \d^{D_4} p_4\, 
\cP(p_5) f(p_4, p_5) \equiv 
\labl{dimreg} \\  & \ 
\frac {1}{2\gp i} \int_{\ominus} \d p_5\, \int_0^\infty \d p_4 
\, \cR_{4}( p_4) \cR_{5}(p_5) \, 
\cP(p_5) \, f(p_4, p_5).
\non 
}
We have introduced the regulator functions $\cR_{4}(p_4)$ and 
$\cR_5(p_5)$ for the 4 dimensional and 5 dimensional integrations, given by
\equ{
\cR_{4}(p_4) =  
\frac {2 \gp^{\half (D_4)}}{\gG(\half D_4)}\, p_4^3 \, 
\Bigl( \frac {p_4}{\gm_4} \Bigr)^{D_4 -4}, 
\qquad 
\cR_{5}(p_5) =  
\frac {\gp^{\half (D_5)}}{\gG(\half D_5)}\, 
\Bigl( \frac {p_5}{\gm_5} \Bigr)^{D_5 -1},
\labl{regufunctions}
}
respectively. 
Here $D_4, D_5$ are the complex extended dimensions for 
the Minkowski and compact space, respectively. The parameters  
$\gm_4, \gm_5$ are renormalization scales. 

For $\Im p_5 > 0$, the pole functions may be written as 
\equ{
\cP^{\ga\ga} = \half 
\Bigl(
- \frac i2 \gp R + \frac \ga{p_5} 
- \gr_-(p_5) 
\Bigr),
~
\cP^{\ga-\ga} = \half 
\Bigl(
- \frac i2 \gp R + \gr_+(p_5)
\Bigr),
\labl{polefunS22>}
}
with 
\equ{
\gr_\ga(p_5) = i\gp R \frac {e^{i\gp R\, p_5}}{1 + \ga e^{i\gp R \, p_5}}.
\labl{finitepolefunS22}
}
For $\Im p_5 < 0$  one obtains
\equ{
\cP^{\ga\ga} = \half 
\Bigl(
 \frac i2 \gp R + \frac \ga{p_5} + \gr_-(-p_5)
\Bigr),
~
\cP^{\ga-\ga} = \half 
\Bigl(
 \frac i2 \gp R - \gr_+(-p_5)
\Bigr). 
\labl{polefunS22<}
}
When these functions  are integrated, using dimensional regularization for 
both the 4 dimensional integral and the KK sum, over the half plane 
contours for which either $\Im p_5 > 0$ or $\Im p_5 < 0$, then we 
can identify their UV behavior. The first terms in \eqref{polefunS22<}
may give rise to genuine 
5 dimensional divergences, the $\alpha/p_5$ term  accounts for 4 dimensional 
divergences, while the remaining terms give the finite contributions.


\begin{thebibliography}{99}

\bibitem{Antoniadis}
I.~Antoniadis, S.~Dimopoulos, A.~Pomarol and M.~Quiros,
Nucl.\ Phys.\ B {\bf 544} (1999) 503
[hep-ph/9810410].

\bibitem{Delgado}
A.~Delgado, A.~Pomarol and M.~Quiros,
Phys.\ Rev.\ D {\bf 60} (1999) 095008
[hep-ph/9812489].

\bibitem{Hall}
N.~Arkani-Hamed, L.~Hall, Y.~Nomura, D.~Smith and N.~Weiner,
[hep-ph/0102090].

\bibitem{Barbieri}
R.~Barbieri, L.J.~Hall and Y.~Nomura,
Phys.\ Rev.\ D {\bf 63} (2001) 105007
[hep-ph/0011311].

\bibitem{Quiros}
A.~Delgado and M.~Quiros,
[hep-ph/0103058].

\bibitem{SHP}
S.~Groot Nibbelink and H.P.~Nilles, work in progress. 

\bibitem{Alvarez}
L.~Alvarez-Gaume, S.~Della Pietra and G.~Moore,
Annals Phys.\  {\bf 163} (1985) 288.

\bibitem{Coleman}
S.~Coleman and B.~Grossman,
Nucl.\ Phys.\ B {\bf 203} (1982) 205.

\bibitem{Alvarez2}
L.~Alvarez-Gaume,
HUTP-85/A092
{\it Lectures given at Int. School on 
Mathematical Physics, Erice, Italy, Jul 1-14, 1985}.

\bibitem{Fischler}
W.~Fischler, H.~P.~Nilles, J.~Polchinski, S.~Raby and L.~Susskind,
Phys.\ Rev.\ Lett.\  {\bf 47} (1981) 757.

\bibitem{Gunaydin}
M.~Gunaydin, G.~Sierra and P.~K.~Townsend,
Nucl.\ Phys.\ B {\bf 253} (1985) 573. \\
M.~Gunaydin, G.~Sierra and P.~K.~Townsend,
Nucl.\ Phys.\ B {\bf 242} (1984) 244.

\bibitem{Pomarol}
A.~Pomarol and M.~Quiros,
Phys.\ Lett.\ B {\bf 438} (1998) 255
[hep-ph/9806263].

\bibitem{Mirabelli}
E.~A.~Mirabelli and M.~E.~Peskin,
Phys.\ Rev.\ D {\bf 58} (1998) 065002
[hep-th/9712214].

\bibitem{deWit}
B.~de Wit, P.~G.~Lauwers and A.~Van Proeyen,
Nucl.\ Phys.\ B {\bf 255} (1985) 569.

\bibitem{Zucker}
M.~Zucker, Off-shell supergravity in five dimensions and 
supersymmetric brane world scenarios, PhD thesis Bonn University, 
BONN-IR-2000-10, ISSN-017208741 \\ 
M.~Zucker,
JHEP {\bf 0008} (2000) 016
[hep-th/9909144].

\bibitem{Bergshoeff}
E.~Bergshoeff, R.~Kallosh and A.~Van Proeyen,
JHEP {\bf 0010} (2000) 033
[hep-th/0007044].

\bibitem{sgn}
S.~Groot Nibbelink, 
hep-th/0108185. 

\bibitem{Capper}
D.~M.~Capper, D.~R.~Jones and P.~van Nieuwenhuizen,
Nucl.\ Phys.\ B {\bf 167} (1980) 479.

\bibitem{GNS}
D.~ Ghilencea, H.P.~ Nilles and S.~ Stieberger, 
hep-th/0108183.


\end{thebibliography}
\end{document}